# A statistical study of sea ice thickness and coverage in the Canadian Arctic

Arya Kimiaghalam[1]

[1]*Department of Physics, University of Toronto, 60 St. George Street, Toronto, ON, M5S 1A7, Canada.*

**The Arctic sea ice cover has significantly declined over the recent decades. The debate on whether this decline is caused by anthropogenic activity or internal cycles is still ongoing. However, despite this uncertainty, some physical factors reinforce this declining trend, one of which is sea ice thickness. The thinning of Arctic sea ice facilitates the melting of sea ice by reducing the heat capacity of the ice volume. The progression of this thinning can potentially accelerate sea ice loss. In this work, we attempt to understand the broad relationship of sea ice cover levels and average sea ice thickness in the Arctic. First, we attempt to understand whether the trend in the Arctic sea ice thickness is statistically significant over multi-year and inter-year seasonal scales, by using mostly non-parametric trend analysis tools. We subsequently study how sea ice thickness, as well as its momentum and fluctuations, are statistically correlated to those of sea ice cover in the Arctic. For this task, we use publicly available Arctic sea ice cover and thickness data from 1979 to 2021, provided by the Pan-Arctic Ice Ocean Modelling and Assimilation System (PIOMAS) and the National Snow and Ice Data Center (NSIDC).**

The overall thinning of the Arctic sea ice has been a topic of interest in atmospheric sciences [1-3]. The recent thinning of Arctic sea ice can potentially facilitate the melting of sea ice by reducing the heat capacity of the ice volume. The progression of this thinning can accelerate sea ice loss. On the other hand, sea ice thickness is difficult to measure, especially on a large scale [4]. Arctic sea ice spans millions of square miles and is constantly displaced by winds and ocean currents. A variety of data sources and analysis techniques are used to study sea ice thickness including satellite data, airborne remote sensing data, modeling (e.g. PIOMAS), and even data observations taken from submarines (e.g. US Navy's Arctic fleet) [4]. The Canadian government has systematically gathered sea ice thickness data from the Northern Territories, largely since 1947, under the Ice Thickness Program [5]. Fig.1 shows the sea ice thickness data from two nearby stations located in the town of Alert, Nunavut. Measurements are taken on a weekly basis, starting after freeze-up, when the ice is safe to walk on. Measurements are made up until the break-up conditions make the ice unsafe [5].

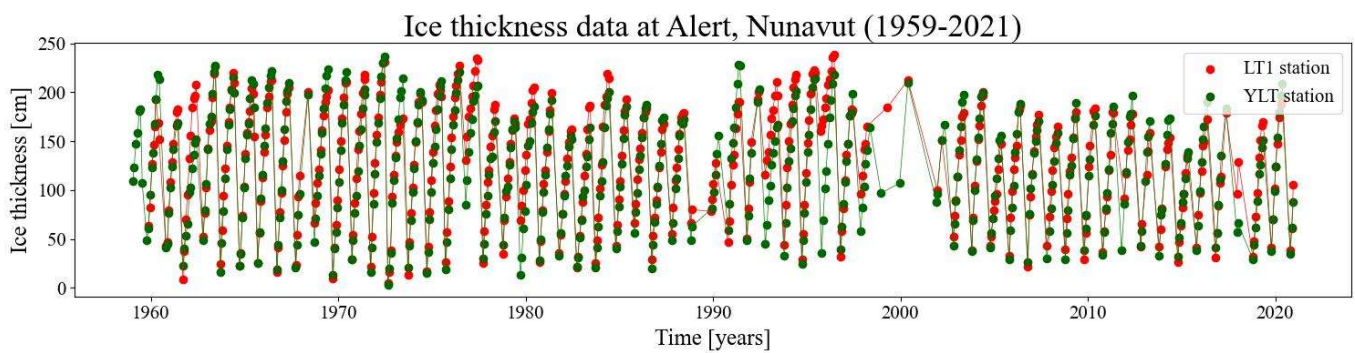

**Fig.1 | Monthly ice thickness data from 1959 to 2021, at LT1 and YLT stations in the town of Alert, Territory of Nunavut, Canada** [5]. The data shows that ice thickness has declined by 0.327 cm/year on average. This figure is substantially lower than the general figure for the Arctic, mainly due to changing ice accumulation patterns and recent growth in ice volume in the Canadian Archipelago region. The original data has been collected on irregular daily intervals. Large gaps are abundant in the Canadian data, best seen in the year 2000. Due to these inconsistencies and irregular periodicity in data collection, analyzing monthly trends and power spectra are not feasible. See supplementary materials for the pan-Canadian map of observation stations.

However, unlike sea ice cover and extent data, ice thickness data, though widely available, includes large gaps and inconsistencies similar to those in the Canadian data shown earlier (see Fig.1). This fact makes it very impractical to use for trend and power spectra analysis. The best dataset was collected by the Unified Sea Ice Thickness Climate Data Record (established by the Polar Science Center Applied Physics Laboratory at the University of Washington) [6-7], however, the dataset lacks support and a consistent dataframe, making it very challenging to process. Thus, for this study, we used Arctic sea ice thickness data provided by the PIOMAS (see supplementary materials for a brief discussion on this selection) [8-9]. Contrary to the local nature of the Canadian data, analysis of the PIOMAS data offers a more comprehensive view of the dynamics involved in sea ice thickness fluctuations [8-9].

## Trends in Arctic sea ice thickness

In this section, the results of a comprehensive trend strength and significance analysis are presented. The average monthly sea ice thickness data is shown in Fig.2. The average ice thickness is declining at a rate of 2.165 cm/year. The inter-year rate of decline over maximum and minimum sea ice cover periods (i.e. March and September respectively), are 1.540 cm/year and 2.699 cm/year respectively.

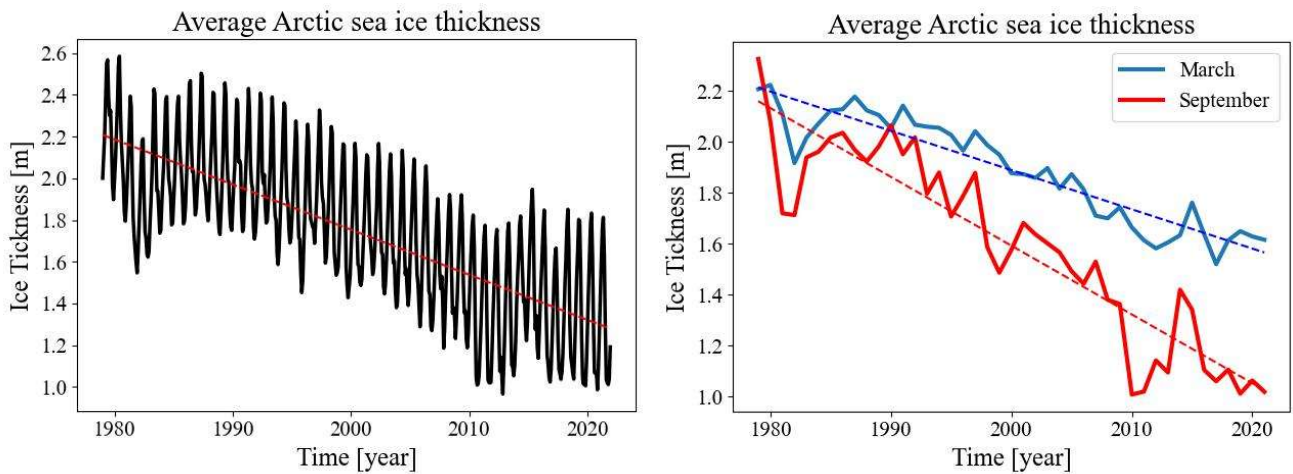

**Fig.2 | Monthly average Arctic sea ice thickness (left), inter-year sea ice thickness in March and September (right).** Trends in the overall, March and September data are 2.165, 1.543 and 2.699 cm/year respectively. All the above trends are statistically significant.

These are very strong trends, however, their significance needs to be statistically examined. To examine trend significance, we use two methods: Monte Carlo reSampling and the Yue&Wang corrected Mann-Kendall test (refer to the Methods section for a detailed discussion of these techniques). The trend in the multi-year monthly sea ice thickness data as well as those over the months of March and September passed both trend significance tests. Therefore, it can be concluded that statistically, the sea ice thickness is significantly declining in the Arctic. This motivates us to look into this data more closely. The ice thickness in September is declining much more rapidly than that of March, potentially implying that the Arctic experiences progressively warmer summers or that conditions favouring the thinning of ice were strengthened. The compiled year-over-year ice thickness data (1979-2021) in Fig.3 offers another interesting clue about these conditions. Since 1979, the boundary of favourable conditions for sea ice accumulation (i.e. growth in ice thickness) has been shifted towards the early months of the year by almost 30 days. On the other hand, the bottom ice levels have not shifted significantly. In other words, the time period from peak to bottom sea ice thickness in the Arctic has been steadily prolonged since 1979. In conclusion, Fig.2 and Fig.3 convey that conditions favouring the thinning of sea ice have simultaneously strengthened and prolonged in time, which is also reflected in literature [2].

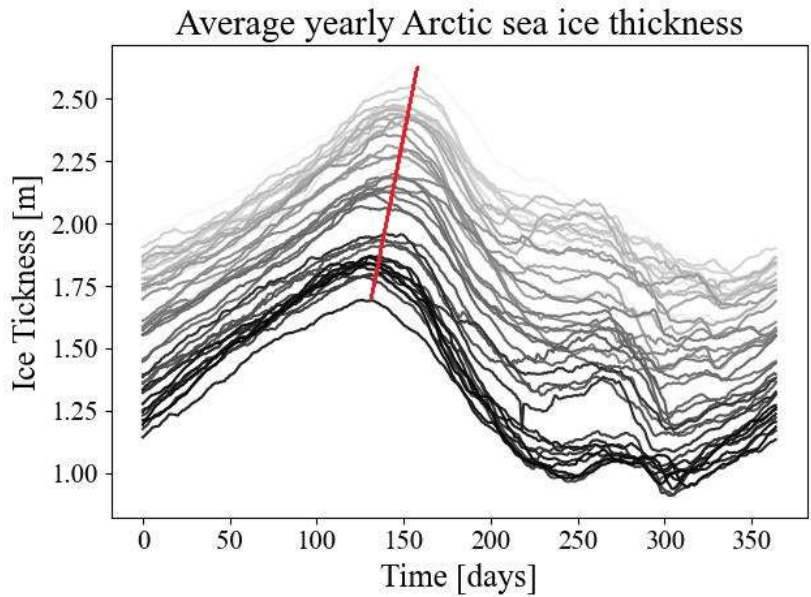

**Fig.3 | Average yearly Arctic sea ice thickness for years 1979 to 2021.** The time series for each year is labeled by a grayscale colour, becoming darker as the years progress forward, with the lightest and darkest being ice thickness in 1979 and 2021 respectively. It can be observed that the ice thickness peaks sometime between early April and mid May, after the sea ice cover peaks. The key observation is the peak positions over the years. The peaks significantly shifted to the earlier months of the year, almost by 30 days (indicated by a solid red line). This potentially indicates that the period of favourable conditions for ice accumulation has shrunk at least by the same amount.

Lastly, we examined if the rate of change in ice thickness has accelerated over this period. We calculated the monthly rate of change in sea ice thickness over one month intervals (i.e. rates are in meters/month). Fig.4 shows the monthly rate of change in sea ice thickness from 1979 to 2021. It can be observed that although the thickness levels have declined significantly, the pace at which it happened has barely accelerated (-6.7 $\mu m/month^2$). This trend failed both trend significance tests, implying that the rate of change in sea ice thickness and its cycle have remained stable over time.

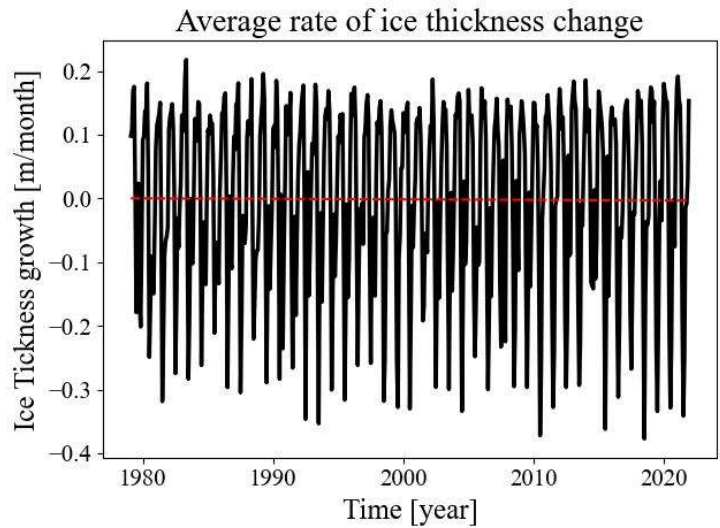

**Fig.4 | Rate of monthly ice thickness growth from 1979 to 2021.** The trend in the data is extremely weak and not statistically significant, implying a fairly stable series of cycles over this period.

# Correlation between Arctic sea ice thickness and cover

In ice dynamics, the rate at which a body of ice melts is a function of temperature gradient, solar radiation, humidity, wind, pressure and salinity. The melting of sea ice influences its physical geometry such as its thickness and cover area above sea level. Note that in this study, an ice covered area is defined by a region that is fully covered by ice, excluding regions that are considered ice extent regions (minimum of 15% ice coverage). Considering that the factors influencing the melting of sea ice affect the physical characteristics of ice simultaneously in time, the change in the physical characteristics of ice are potentially correlated to each other.

The most simple system that involves ice cover and thickness growth is the freezing of a pond. Although this example involves a very different set of dynamics than the Arctic ocean, it can offer us some useful, first order insights about ice coverage and its relationship to thickness. Unlike the vast majority of liquids, water freezes *top-down*. This means that initially, a thin layer of ice forms on the surface (i.e. growing ice cover), and after some time, the growth in thickness starts to take off as the ice cover described earlier acts as an insulator, reducing heat transfer between the above-surface environment and the deeper waters, further facilitating ice formation. This lag between the change in ice cover and thickness may further complicate the correlation between these two factors. In this study, we first examine the correlation of Arctic sea ice cover and thickness over a yearly cycle. Fig.5 shows the correlation between average ice thickness and cover for the years 1979 to 2021.

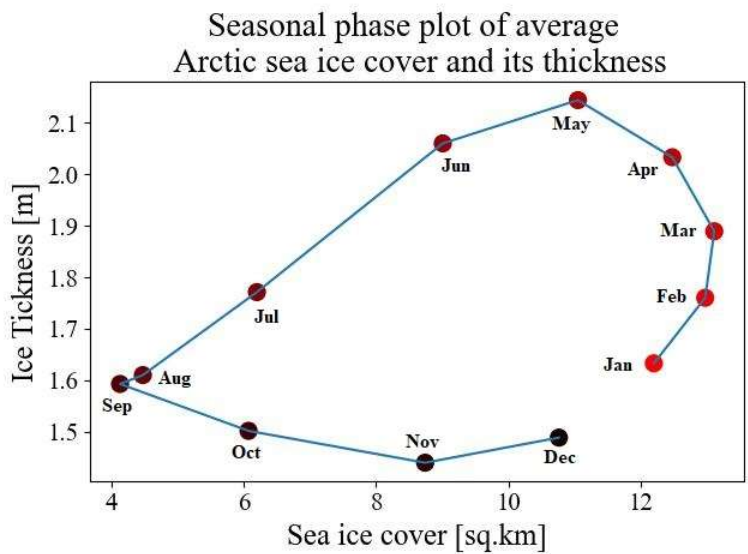

**Fig.5 | Average Arctic sea ice thickness vs. average ice cover for the months from 1979 to 2021**. Data points belonging to each month have been labeled accordingly. The general elliptical shape of the cycle remains. Interpreting this ellipse-shaped cycle is a difficult task. However, a likely conclusion is that Arctic sea ice cover and thickness have a relative phase shift in how they take effect, assuming a similar frequency profile for sea ice thickness and cover (dominated by the yearly frequency of seasons). In addition, either sea ice cover or thickness data could contain a higher order periodicity, such as an $n^{th}$ power of a sinusoidal function. The addition of such higher power oscillations can potentially explain the disuniform deformation of the ellipse (Aug-Oct vs. Feb-Apr).

The behaviour of the cycle in Fig.5 shows that the average yearly cycle in Arctic sea ice thickness and cover are most likely not directly proportionate. Two directly proportional cyclic events would have a linear phase plot. Introducing a phase shift between two proportional cycles produces an elliptically-shaped phase plot similar to that in Fig.5. The following describes the so-called phase shift described earlier: starting from November, the sea ice cover and thickness begin to grow. The growth in both thickness and cover persists until March, where the trend in sea ice cover growth reverses. Nonetheless, the ice thickness keeps growing by another two months, peaking in May. The next period of similar shift happens after August, where the sea ice cover gradually starts to recover from its lows. However, despite the growth in ice cover, the ice keeps thinning, though very gradually, until November. Another interesting observation is how the supposed ellipse has been differently deformed during each part of the cover-thickness cycle. For instance, we can consider the movement in the phase plot for March-to-May and September-

November periods. These are equal time intervals between maximum sea ice cover and extent, and minimum sea ice cover and extent respectively. The average rate of change in sea ice thickness in time is much more rapid during the transition from maximum cover and extent (March-May), compared to the transition between the minima (September-November). The introduction of an $n^{th}$ order oscillatory function (e.g. $sin^2(wt)$) to either sea ice cover or thickness evolution can closely model the unique asymmetry in Fig.5. However, finding the exact parameters of this relationship requires an in-depth analysis of sea ice dynamics, which is beyond the scope of this study. Nonetheless, this observation further reinforces the idea that ice thickness growth is a complex function of ice cover. On the other hand, if the two cycles (in this case sea ice cover and thickness) have significant differences in their frequency profile, the shape departs from being a closed ellipse, to a complex, self intersecting curve. This prompts us to study the frequency profile of the monthly sea ice thickness and cover in the next section.

### Power spectra of sea ice thickness and cover

In addition to the full monthly data from 1979 to 2021, the direct correlations between the momenta in the months of September and March, as well as those for ice thickness and cover fluctuations were also studied. However, the direct correlation results did not indicate any strong correlation between September and March ice thickness, cover and their respective momentum data ($p < 0.5$). The only exception was the correlation between September sea ice cover and ice thickness, with a Pearson correlation of 0.813. However, the correlation in their fluctuations is only 0.429. This implies that the sheer trends in sea ice cover and thickness makes up most of their correlation; effectively conveying that the long term pace of ice cover decline is implicitly linked to the thinning of the Arctic ice body. However, due to the cyclicity of Arctic sea ice parameters, a better way to find similarities between the monthly sea ice cover and thickness data is through their power spectra. Power spectra of time series are great tools to analyze their frequency composition. As briefly discussed previously, Fig.5 may imply that the frequency components of ice thickness and cover have to be fairly identical; although, each of those frequencies are associated with different strengths (i.e. power). Fig.6 shows the power spectra for sea ice thickness, cover, and their respective momentum time series. Here, momentum is simply defined as the rate of change in a time series.

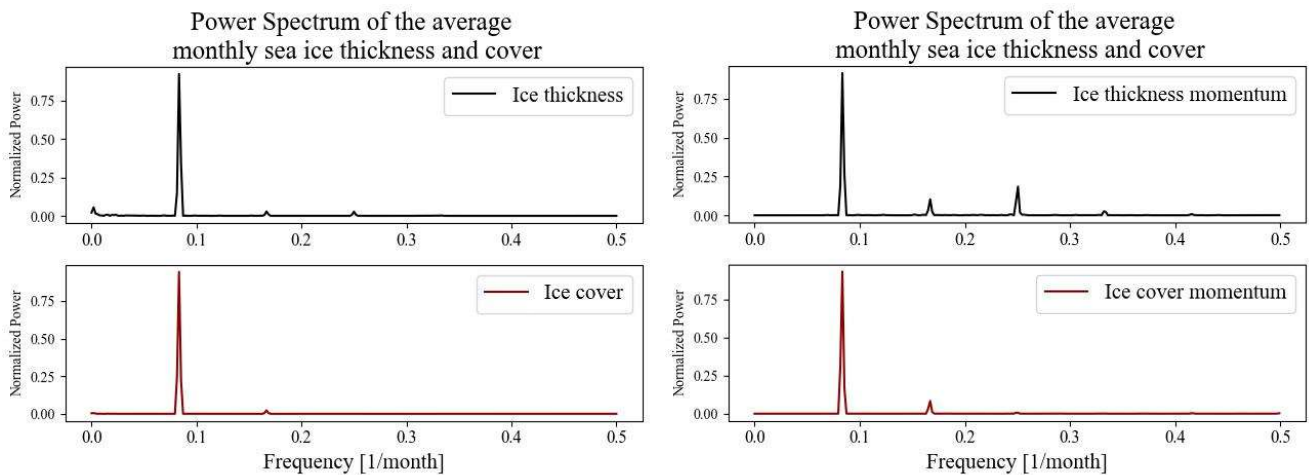

**Fig.6 | Normalized power spectrum of sea ice thickness, cover (left) and their momenta (right).** All time series, including those of momenta, share 12 months and 6 months distinct frequency components. However, unlike ice cover, the ice thickness time series (disregarding extremely low frequencies) and its momentum contain a minor 4 months frequency component, with a sizable power amplitude for ice thickness momentum. Note that all the above time series were windowed (by Hanning windowing) prior to the calculation of the spectra, in order to minimize spectral leakage.

We can observe that the ice thickness and cover time series, as well as their momenta, share a 12 months frequency component, with the 12 months period (i.e. the yearly frequency) strongly dominating other components. Surprisingly, the power spectra for sea ice thickness and its momentum contain 4 and 6 month frequency components, with the 4 month period having a significantly large amplitude for the sea ice thickness momentum time series. This means that the rate of change in sea ice thickness is significantly influenced by 4 and 6 month periodic drivers. Lastly, due to the relatively small amplitudes of the 4 and 6 months periods, the possibility of spectral leakage or the effect of aliasing were examined. A Hanning window was applied to the data and the power spectra were re-calculated. However, the 4 and 6 month periods persisted, adding confidence to the results in Fig.6.

Isolating a set of physical factors that contribute to the 4 month period component of sea ice thickness momentum is extremely difficult, given that climate phenomena with 4 month periods are extremely rare. In addition, other distorting factors such as non-stationarity of underlying frequencies can further complicate our power spectrum analysis. Nonetheless, these results surely motivate further studies in the physical dynamics of sea ice thickness growth and fluctuations in the Arctic.

## Discussions and Conclusion

In conclusion, using non-parametric statistical tools, such as the Yue&Wang Mann-Kendall test and Monte Carlo resampling, we identified a statistically significant declining trend in the Arctic sea ice thickness data. In addition, through analyzing the compiled monthly sea ice thickness data (Fig.3), we discovered that the conditions favouring the thinning of sea ice have been strengthened and prolonged from 1979 to 2021. Through further analysis of the Arctic sea ice cover and thickness data, we showed that the average sea ice cover and thickness generally follow a cyclical pattern. This cycle is prone to being shifted along the *y=x* line in the cover-thickness space over time, however the cyclic nature is persistent. This cyclical behaviour implies a similar frequency profile for sea ice thickness and sea ice cover in the Arctic, with a pronounced relative phase shift, or *"time delay"* in their underlying dynamics. Lastly, we calculated the power spectra of sea ice cover, sea ice thickness, and their momenta. We discovered that in addition to a very strong 12-month (i.e. yearly) periodicity, a 4 and 6-month periodicity was present in ice cover and thickness data, with the 4-month periodicity being fairly significant in sea ice thickness and its momentum. These periodicities are extremely rare in atmospheric and planetary dynamics, motivating further analysis and work on the physics behind the dynamics of large ice bodies.

# Methods

## *Yue and Wang corrected MK trend test*

A method for estimating trend significance is the Mann-Kendall (MK) test. The test involves calculating the difference between all pairs of data points and determining the sign of each difference. This is called the MK statistic:

$$S = \sum_{i=1}^{n-1} \sum_{j=i+1}^{n} \text{sgn}(X_j - X_i)$$

Where the $X_j$ are the sequential data values, n is the length of the data set, and:

$$\text{sgn}(\theta) = \begin{cases} 1 & if\ \theta > 0 \\ 0 & if\ \theta = 0 \\ -1 & if\ \theta < 0 \end{cases}$$

A trend is detected if the number of positive differences is significantly different from the number of negative differences [10]. The test is robust to non-normality and outliers in the data, making it a useful tool for trend analysis in environmental and climatological studies. By calculating the variance and knowing the MK statistic, we can calculate the P-value associated with our trend:

$$V(S) = \frac{n(n-1)(2n+5) - \sum_{i=1}^{n} t_i\, i\, (i-1)(2i+5)}{18}$$

and

$$Z = \begin{cases} \dfrac{S-1}{\sqrt{Var(S)}} & S > 0 \\ 0 & S = 0 \\ \dfrac{S+1}{\sqrt{Var(S)}} & S < 0 \end{cases}$$

$$p = 0.5 - \Phi(|Z|)$$

$$(\Phi(|Z|) = \frac{1}{\sqrt{2\pi}} \int_0^{|Z|} e^{-\frac{t^2}{2}} dt)$$

Where P is the p-value and Var(S) = V(S).

However, the test produces subpar significance estimations for highly autocorrelated data. Yue and Wang [10] (YW) propose a modified version of the MK test which takes the data's autocorrelation and reduced degrees of freedom into account. The modification involves a pre-whitening procedure which fits an autoregressive AR(1) model to the data and computes the residuals [10-14]. The AR(1) model captures the lag-1 autocorrelation in the data and provides an estimate of the data's autocorrelation structure. The residuals are then used in place of the original data in the MK test, and the resulting statistic is adjusted using an adjustment factor to account for the reduction in sample size due to this procedure [10-14]. The YW correction is implemented through the `pymannkendall` python library.

## *Monte Carlo reSampling Technique*

The Monte Carlo resampling technique is implemented by randomly generating a number of time series (O($10^5$) in this case) that share similar characteristics to the time series under investigation, and then estimate the p-value of the trend in the original time series based on the probability density distribution of these generated time series' trend values. However, there can be several parameters that two time series could share. In this implementation of the test, our generated time series share the same power spectrum as the time series under investigation. To generate such time series each time, we first Fourier transform the time series under investigation, using a Fast Fourier Transform (FFT) algorithm. Then, we re-assign random phases to its Fourier coefficients, and inverse transform it back into real space [15]; and this process is repeated many times ($10^5$ times in this implementation).

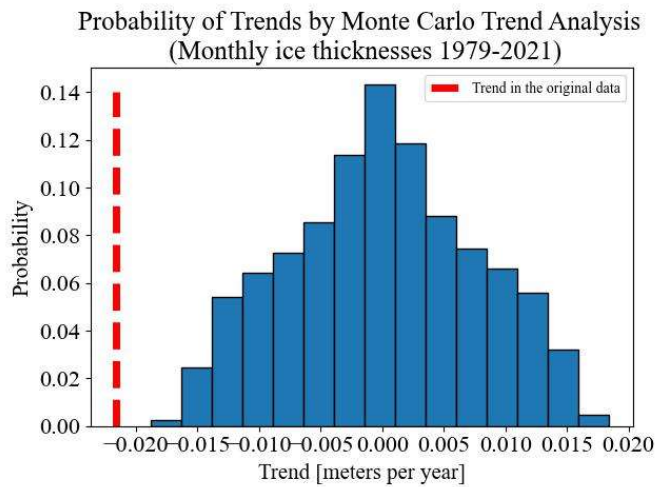
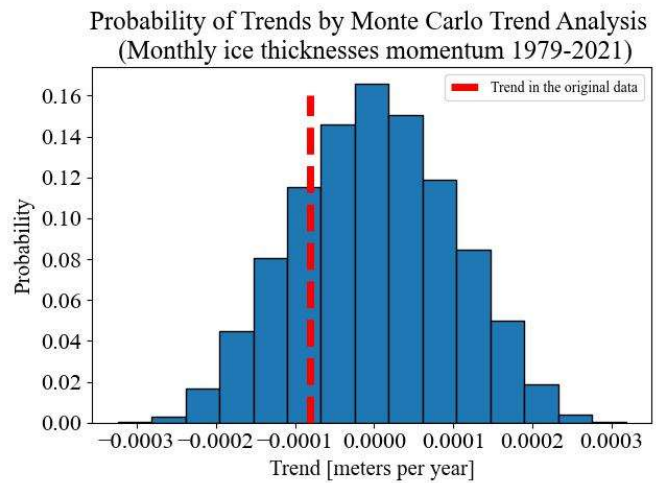
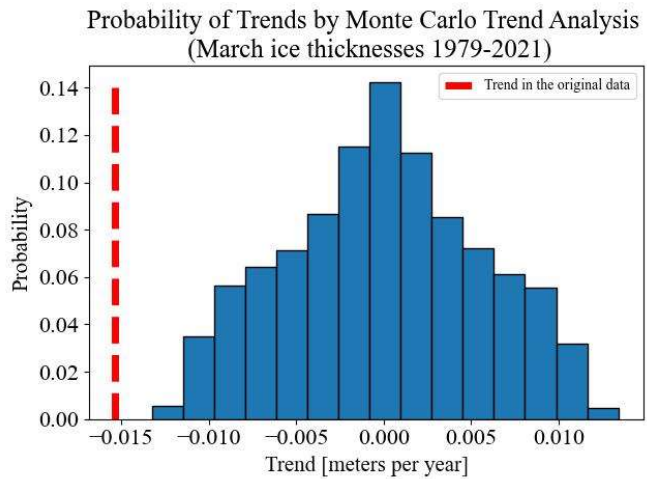
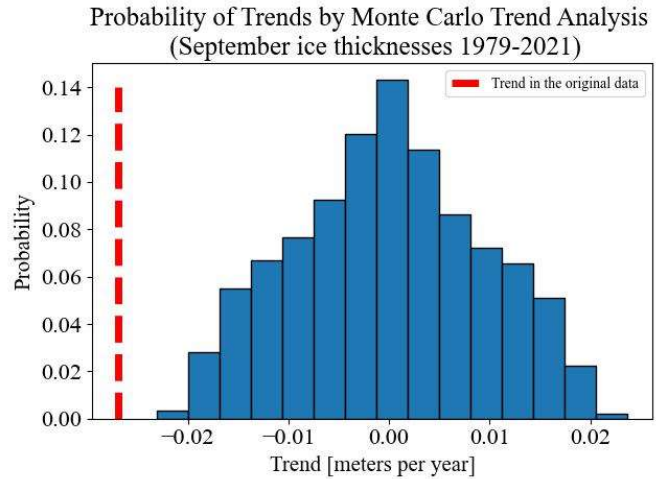

**Fig.A | Probability distributions of trends in the monthly Arctic sea ice thickness data (overall data as well as that for the months of March and September, which are the months of maximum and minimum of Arctic sea ice cover), as well as its momentum time series.** The trend in the original data is marked by red dashed lines. All of these figures indicate that the trends in the ice thickness data are statistically significant ($p < 0.05$). The fat-tail nature of these distributions are an indication of the red noise characteristic of the sea ice thickness data (and geophysical data in general).

This approach preserves the autocorrelated nature (i.e. red noise nature) of the original time series and helps us to better identify the trend significance for highly correlated time series, such as sea ice data [15]. Fig.A shows the trend distribution over $10^5$ iterations of the Monte Carlo resampling, implemented for Arctic sea ice thickness and its momentum. Note that due to the stochasticity of this approach, the probability density distributions can change over different runs. Thus, only a rough estimate of the trend significance can be obtained.

# **Supplementary Materials**

## *Map of measurement stations of the Canadian Ice Thickness Program (1947-2021)*

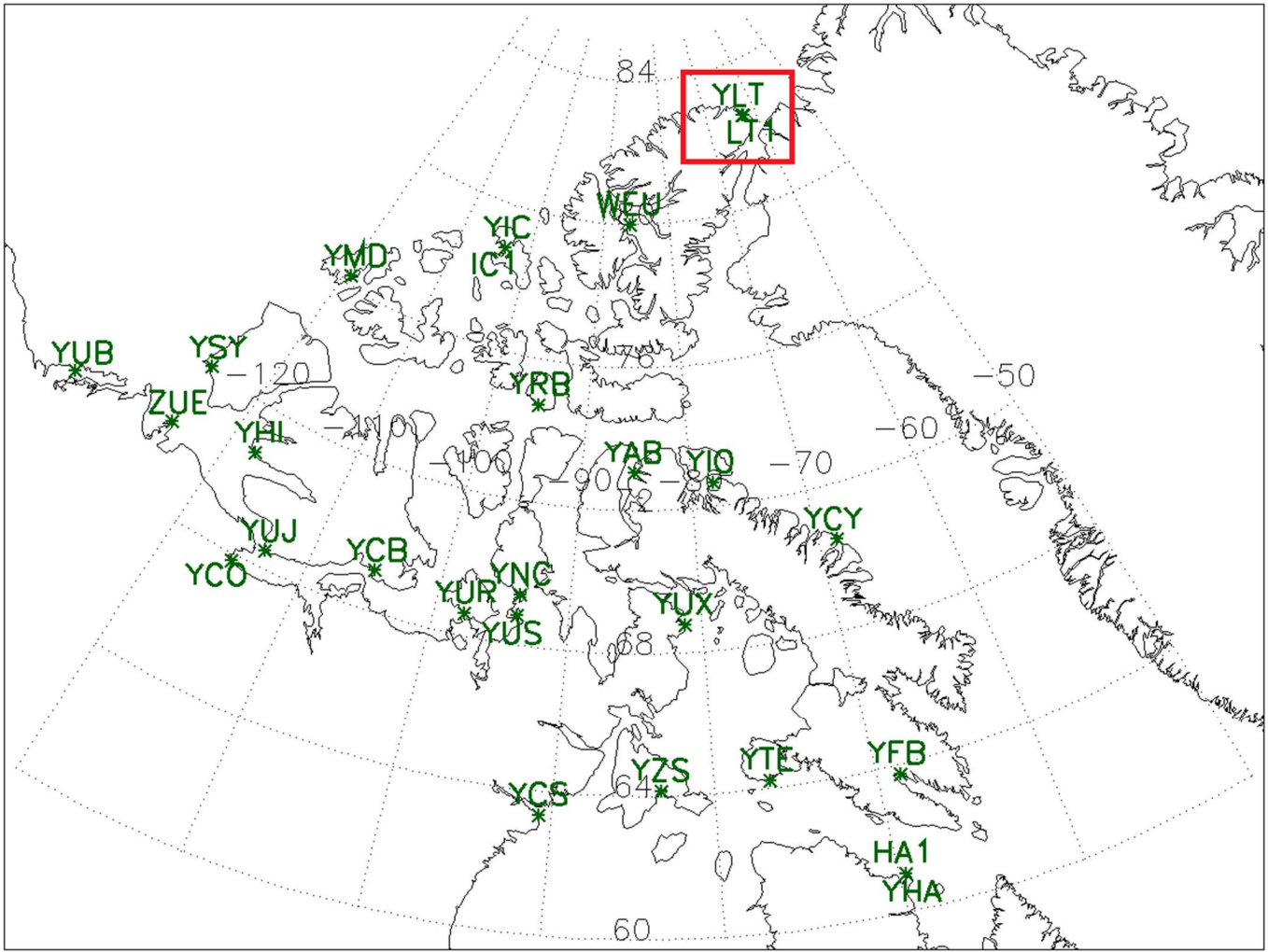

**Fig.B | Map of ice thickness measurement stations operating under the pan-Canadian ice thickness program. All stations gathered information from 1947 to 2003.** However, only the following stations collected data from 2003 to 2021: LT1, YLT, YBK, YCB, YZS, WEU, YUX, YEV, YFB, YRB and YZF [5].

*Note that the data from the stations labeled in the red box above are used for plotting Fig.1.*

## *Justification for using PIOMAS ice thickness data together with NSIDC ice cover data*

We coupled PIOMAS average Arctic sea ice thickness data with NSIDC ice cover data. They are based on climate models and direct observation respectively. This choice was made mainly due to the disparity in the completeness of the data for ice thickness compared to ice cover in publicly available data [6]. The first question that could be raised is the compatibility between simulated and directly observed data. We will examine this compatibility through assessing the influence of this choice on our analysis of the ice thickness-cover cycle (see Fig.5) as well as our power spectrum analysis of the ice thickness and cover data (see Fig.6).

The fundamental cyclicity in Fig.5 is affected by the frequency and phase composition. Fig.C shows the PIOMAS and NSIDC Arctic sea ice cover data from 1979 to 2021. Although PIOMAS ice cover has not been directly used in this study, we can use it as a representation of the PIOMAS simulation, since both PIOMAS sea ice thickness and cover are the outcome of the same dynamic climate model.

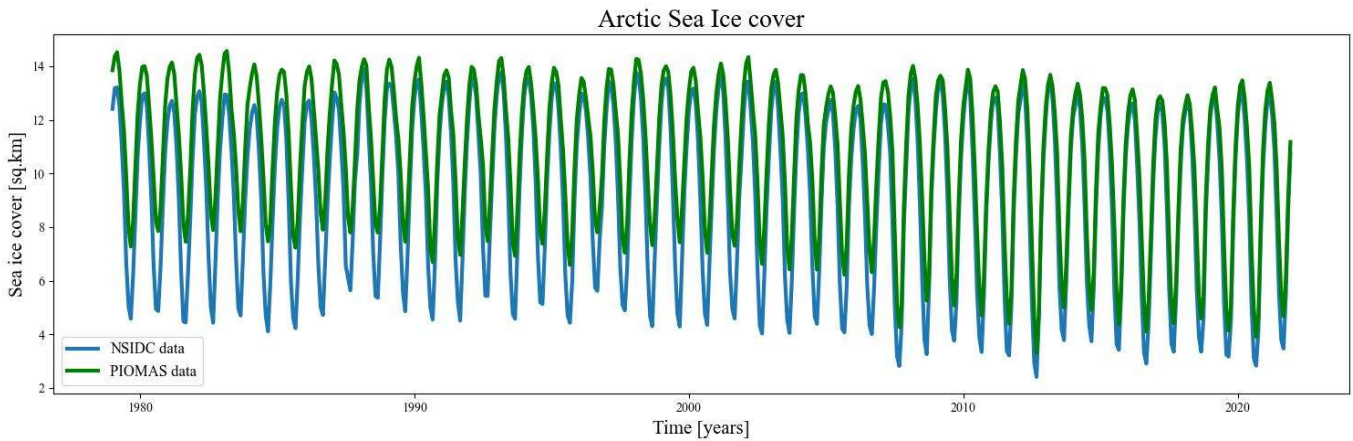

**Fig.C | The monthly Arctic sea ice cover data from the NSIDC (in blue) and PIOMAS (in green) from 1979 to 2021.** Although these time series differ in amplitude and mean, they are in tune periodically as well as in their phase. Sea ice covers are in km$^2$ and time is in years.

We can observe that although these time series have different amplitudes and means over time, they are in-tune in frequency and amplitude. This means that switching between the use of NSIDC or PIOMAS data to generate Fig.5 would only change the shape of the elliptical cycle (through vertical and horizontal stretching) and not undermine the closed-loop, cyclic nature of the ice cover-thickness phenomenon. Thus, we can have more confidence in simultaneously using direct observation of sea ice cover and simulated sea ice thickness in our study. We can further strengthen this argument by analyzing the power spectra of the time series in Fig.C (See Fig.D).

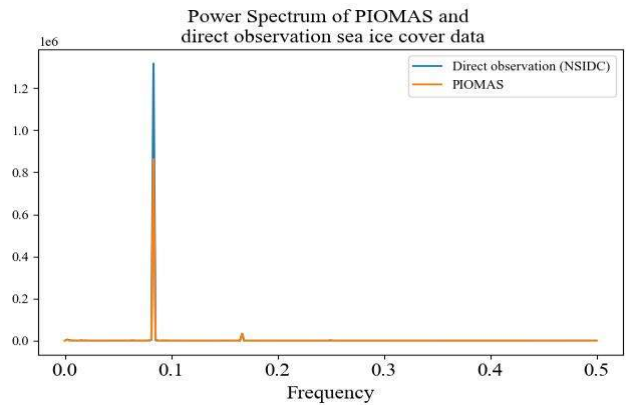

**Fig.D | Power spectra of the PIOMAS and NSIDC (i.e. directly observed) sea ice cover data from 1979 to 2021.** The spectra show identical peak positions. Both contain the 12 and 6 month periodicity. Frequency is reported in month$^{-1}$ (The power spectra are not normalized).

## Acknowledgment

I would like to thank George William Kent Moore for offering me advice and helpful discussions.